\documentclass[prod]{jfmfr}
\usepackage{epsfig}
\usepackage[english]{babel}
\usepackage{textcomp}
\usepackage{amsmath}
\usepackage{color}
\usepackage{amssymb}
\usepackage{hyperref}
\usepackage{epic}
\usepackage{pict2e}
\usepackage{epstopdf}
\usepackage{pict2e} 
\usepackage{ulem}

\begin{document}

\shorttitle{Advection and diffusion in a chemically induced compressible flow}
\shortauthor{F. Raynal, M. Bourgoin, C. Cottin-Bizonne, C. Ybert and R. Volk}

\title{Advection and diffusion in a chemically induced compressible flow}

\author{Florence Raynal\aff{1} \corresp{\email{florence.raynal@ec-lyon.fr}},
  Mickael Bourgoin\aff{2},
  C\'ecile Cottin-Bizonne\aff{3},
  Christophe Ybert\aff{3}
 \and Romain Volk\aff{2} \corresp{\email{romain.volk@ens-lyon.fr}}
}
 
\affiliation{\aff{1}LMFA, Univ Lyon, \'Ecole Centrale Lyon, INSA Lyon, Universit\'e Lyon 1, CNRS, F-69134 \'Ecully, France. \aff{2}Laboratoire de Physique, ENS de Lyon, Univ Lyon, CNRS, 69364 Lyon CEDEX 07, France. \aff{3}Institut Lumi\`ere Mati\`ere, Univ Lyon, Universit\'e Lyon 1, CNRS, F-69622 Villeurbanne CEDEX, France.}

\date{28 January 2018; revised 5 April 2018; accepted 13 April 2018} %--> renseigné dans le fichier jfmfr.cls
\pubyear{2018} % only needed when year of publication is not current year.
\volume{847}
\pagerange{228--243}
\doi{10.1017/jfm.2018.335}

\maketitle

\begin{abstract}

We study analytically the joint dispersion of Gaussian patches of salt and colloids in linear flows, and how salt gradients
accelerate or delay colloid spreading by diffusiophoretic effects. Because these flows have constant gradients in space, the problem can be solved almost entirely for any set of parameters, leading to predictions of how the mixing time and the Batchelor scale are modified by diffusiophoresis. We observe that the evolution of global concentrations, defined as the inverse of the patches areas, are very similar to those obtained experimentally in chaotic advection. They are quantitatively explained by examining the area dilatation factor, in which diffusive and diffusiophoretic effects are shown to be additive and appear as the divergence of a diffusive contribution or of a drift velocity. An analysis based on compressibility is developed in the salt-attracting case, for which colloids are first compressed before dispersion, to predict the maximal colloid concentration as a function of the parameters. This maximum is found not to depend on the flow stretching rate nor on its topology (strain or shear flow), but only on the characteristics of salt and colloids (diffusion coefficients and diffusiophoretic constant) and the initial size of the patches. 

\end{abstract}


\section{Introduction}

The transport of colloidal particles by a flow can be greatly modified by the presence of a scalar in the fluid. In the case of an electrolyte, (salt) concentration gradients are at the origin of diffusiophoresis which results in a drift velocity  $\mathbf{v}_\mathrm{dp} = D_\mathrm{dp} \nabla \log S$ between the colloids and the flow, where $S$ and $D_\mathrm{dp}$ are the salt concentration and the diffusiophoretic coefficient (\cite{bib:Anderson1989,bib:Abecassisetal2009}).

Recent experimental and numerical studies showed how the mixing of colloids undergoing chaotic advection is strongly modified by diffusiophoresis. In particular, \cite{bib:Deseigneetal2014} showed that the time needed to mix the colloids can be strongly increased or decreased depending on whether salt and colloids are released together or not, an effect further explained based on the compressibility of the drift velocity (\cite{bib:Volketal2014}). 
Investigating gradients of the colloid concentration field in more recent experiments, \cite{bib:Maugeretal2016} demonstrated diffusiophoresis acts both at large and small scales, resulting in a modification of the Batchelor scale at which stretching and diffusion balance (\cite{bib:batchelor1959}). 
All these observations were made in a limited range of P\'eclet numbers, and no general prediction was made concerning diffusiophoresis in the limit of large stretching, or vanishingly small colloid diffusion coefficient. In this article, we obtain analytical predictions on the dispersion of two-dimensional (2-D) patches of salt and colloids advected by linear velocity fields (pure deformation and pure shear) which are chosen to correspond to fundamental examples at the heart of our understanding of mixing (\cite{bib:Townsend1951,bib:Taylor1953,bib:batchelor1959,bib:bakuninbook}). 
Diffusion and diffusiophoresis are examined in light of the area dilation factor of the patches,  in which diffusive and diffusiophoretic effects are shown to be additive and appear as the divergence of a diffusive contribution or of a drift velocity.
This allows for a quantitative prediction of the maximum concentration of the colloids in the salt-attracting case, for which colloids are first compressed before dispersion. This maximum is found not to depend on the flow stretching rate nor its topology (strain or shear flow), but only on the characteristics of salt and colloids (diffusion coefficients and diffusiophoretic constant) and the initial size of the patches.

In the presence of diffusiophoresis, the salt and colloids with respective concentrations $S(x,y,t)$ and $C(x,y,t)$ evolve following the set of coupled advection--diffusion equations:
\begin{eqnarray}
&\frac{\displaystyle \partial S}{\displaystyle \partial t} + \nabla \cdot  [S \mathbf{v}] = D_s\, \nabla^2 S,\label{eqs}\\
&\frac{\displaystyle \partial C}{\displaystyle \partial t} + \nabla \cdot [C (\mathbf{v}+ \mathbf{v}_{\mathrm{dp}})] = D_c\, \nabla^2 C,
\label{eqc}\\
&\mathbf{v}_{\mathrm{dp}} = D_{\mathrm{dp}}\; \nabla\log S
\label{eq:vdp},
\end{eqnarray}
%where $\mathbf{v}$ is a 2D divergence free velocity field, and $D_s$, $D_c$, $D_\text{dp}$ are respectively the diffusion coefficients of both species and the diffusiophoretic coefficient (\cite{bib:Deseigneetal2014,bib:Volketal2014,bib:Maugeretal2016}). 
where $\mathbf{v}$ is a divergence free velocity field, and $D_s$, $D_c$, $D_\mathrm{dp}$ are respectively the diffusion coefficients of both species and the diffusiophoretic coefficient (\cite{bib:Deseigneetal2014,bib:Volketal2014,bib:Maugeretal2016}). 
Here we take $\mathbf{v} = (\sigma x,-\sigma y)$ (pure deformation),  or $\mathbf{v} = (\gamma y,0)$ (pure shear).
When the patches of colloids and salt are released at the origin with initial sizes $y_{0,c}$ and $y_{0,s}$, their dispersion is governed by the colloids P\'eclet number $Pe_c = \sigma y_{0,c}^2 / D_c$, the salt P\'eclet number $Pe_s = \sigma y_{0,s}^2 / D_s$, and the diffusiophoretic number $D_\mathrm{dp}/D_s$, where $\sigma$ would be replaced by $\gamma$ when considering the case of shear instead of deformation.

As the velocity fields we consider are linear, patches of salt released with Gaussian profiles will remain Gaussian at all times (\cite{bib:Townsend1951,bib:batchelor1959,bib:bakuninbook}), resulting in a drift velocity $\mathbf{v}_{\mathrm{dp}} = D_{\mathrm{dp}}\; \nabla\log S$ which is also a linear flow. Assuming the colloids are released with a Gaussian profile too, both cases of deformation and shear can be analytically solved almost entirely, leading to predictions in the limit of large P\'eclet numbers.

The article is divided as follows: section \ref{puredef} is devoted to the evolution of the patches under pure deformation in the velocity field $\mathbf{v} = (\sigma x,-\sigma y)$, which corresponds to a stagnation point. 
This example is solved analytically using the method of moments, and allows for a prediction of the Batchelor scale for the colloids $\ell_{B,c} = \sqrt{D_cD_s/\sigma (D_\mathrm{dp} + D_s)}$ at which occurs an equilibrium between advection and diffusion (\cite{bib:batchelor1959}). 
Section \ref{pureshear} deals with the more complex case of a pure shear flow $\mathbf{v} = (\gamma y,0)$, that can be solved numerically for the colloids with any set of parameters. 
Section \ref{mixcompress} is the heart of the article where it is shown that the present results, although corresponding to academic cases, are very similar to those obtained experimentally. This section discusses the time evolution of the colloid concentration using arguments based on compressibility, and allows for a prediction of i) how the mixing time varies in the limit of small diffusivities with or without diffusiophoresis. ii) what is the maximum concentration that can obtained when salt and colloids are released together, corresponding to a configuration in which the colloid concentration is first amplified before being attenuated. We show here that this maximum (divided by its initial value) is independent of the flow parameters and scales as 
\begin{equation}
\tilde{c}_\mathrm{max}=\displaystyle \left(\frac{D_\mathrm{dp}}{D_c}\right)^{\frac{D_\mathrm{dp}}{2(D_\mathrm{dp}+D_s)}}
\end{equation}
when the patches have the same initial size. 
Finally, section \ref{concl} gives a summary of the various results and explains why these are not modified in the large stretching limit.

\section{Dispersion under pure deformation}
\label{puredef}
\subsection{Initial configuration and notations}

The first problem we address is the joint evolution of 2-D patches of salt and colloids, released together at the origin, under pure deformation by a linear velocity field $\mathbf{v} = (\sigma x, -\sigma y)$ ($\sigma \geq 0$). 
This corresponds to a stagnation point with a dilating direction $(x)$ and a contracting direction ($y$).
Assuming the patches have initial Gaussian profiles  
\begin{eqnarray}
S(x,y,t=0) & = & \frac{N_s}{2 \pi \sqrt{x_{0,s}^2 y_{0,s}^2}}\,\exp\left(-\frac{x^2}{2 x_{0,s}^2}-\frac{y^2}{2 y_{0,s}^2}\right), N_s\geq0, \\
C(x,y,t=0) & = & \frac{N_c}{2 \pi \sqrt{x_{0,c}^2 y_{0,c}^2}}\,\exp\left(-\frac{x^2}{2 x_{0,c}^2}-\frac{y^2}{2 y_{0,c}^2}\right), N_c\geq0, 
\end{eqnarray}
we define the P\'eclet numbers  along the contracting direction ($y$) for both salt $Pe_s = {\sigma y_{0,s}^2}/{D_s}$ and colloid $Pe_c = {\sigma y_{0,c}^2}/{D_c}$.\\

\noindent The different configurations under study in this articles are chosen to correspond to those studied in \cite{bib:Deseigneetal2014,bib:Volketal2014,bib:Maugeretal2016}:\\% satin pas défini dans bib:Abecassisetal2009,
\quad (i) {\it No salt:} $D_\mathrm{dp} =0$ which is the reference case (no diffusiophoresis) and corresponds to simple advection-diffusion of both species.\\
\quad (ii)  {\it Salt attracting:} $D_\mathrm{dp} >0$ which leads to delayed mixing at short time due to the drift velocity $\mathbf{v}_{\mathrm{dp}} = D_\mathrm{dp} \nabla\log S$.\\
\quad (iii) {\it Salt repelling:} $D_\mathrm{dp} <0$ which leads to accelerated mixing at short time due to the drift velocity $\mathbf{v}_{\mathrm{dp}} = D_\mathrm{dp} \nabla\log S$. 
Such situation would be obtained by replacing classical salt with a ionic surfactant such as Sodium Dodecyl Sulfate (SDS) (\cite{bib:Banerjee_etal2016})\footnote{In experiments, the salt-repelling case could also correspond to $D_\mathrm{dp}>0$ and an initial profile of the form $C'=\mathrm{max}(C)-C$, which would have infinite colloid content and variance. However, as the salt concentration field is Gaussian, $\nabla \cdot \mathbf{v}_{\mathrm{dp}}$ is constant in space which would lead to an artificial amplification of the concentration $C$ in this case. We thus chose to treat the salt-repelling case by reversing the sign of $D_\mathrm{dp}$. Note that those two situations led to similar mixing times when tested numerically with flow fields that display chaotic advection.}.

%%%%%%%
\subsection{Case of salt, pure deformation}

As mentioned before, when advected by a linear velocity field, an initially Gaussian profile will remain Gaussian at all times. Introducing the total salt content $\iint_\infty S(x,y,t) dx dy =N_s$, and the moments of the concentration profile
\begin{equation}
\langle x^\alpha y^\beta\rangle_s(t)= \frac{1}{N_s} \iint_\infty x^\alpha y^\beta S(x,y,t) dx dy,
\end{equation}
the salt concentration writes at all times
\begin{equation}
S(x,y,t)=\frac{N_s}{2\pi \sqrt{\Delta_s(t)}}\,\exp\left(-\frac{1}{2\Delta_s(t)}\bigl[\langle y^2\rangle_s\,x^2-2\langle xy\rangle_s\, xy+\langle x^2\rangle_s\, y^2\bigr]\right), \label{gauss_s}\\
\end{equation}
where $\Delta_s(t)=\langle x^2\rangle_s\langle y^2\rangle_s-\langle xy\rangle_s^2$ is related to the area ($\mathcal{A}_s$) of the salt patch by the relation $\mathcal{A}_s = \sqrt{\Delta_s}$. In the case of Gaussian profiles, equation (\ref{eqs}) can be solved by using the method of moments (\cite{bib:Aris56,bib:birchetal2008}). This method transforms the original equation for the concentration $S(x,y,t)$ in a set of Ordinary Differential 
Equations (ODEs) for the moments, which are obtained by taking corresponding moments of equation (\ref{eqs}). 
For the case of Gaussian profiles we consider, only second order moments are needed and the system writes:
\begin{eqnarray}
\frac{d\langle x^2\rangle_s}{dt}-2\sigma \langle x^2\rangle_s &=&2D_s
\label{eq:x^2sel_deformation}\\
\frac{d\langle xy\rangle_s}{dt}-\sigma\langle xy\rangle_s&=&0\\
\frac{d\langle y^2\rangle_s}{dt}+2\sigma\langle y^2\rangle_s &=&2D_s\;,
\label{eq:y^2sel_deformation}
\end{eqnarray}
with initial conditions $\langle x^2\rangle_s(0)=x_{0,s}^2$, $\langle y^2\rangle_s(0)=y_{0,s}^2$ and $\langle xy\rangle_s(0)=0$. The second-order moments form a closed set of ODEs (\cite{bib:younggarrett1982,bib:rhines_young1983}) which has solutions (\cite{bib:bakuninbook}): %\NOTE{young montre que le system est ferm\'e pour le shear, pas pour le point de stagnation, mais je cite quand meme} 
\begin{eqnarray}
\langle x^2\rangle_s(t)&=&\left(x_{0,s}^2+\frac{D_s}{\sigma}\right)\exp(2\sigma t) -\frac{D_s}{\sigma}\,,\label{x^2selfinaldef}\\
\langle y^2\rangle_s(t)&=&\left(y_{0,s}^2-\frac{D_s}{\sigma}\right)\exp(-2\sigma t) +\frac{D_s}{\sigma}\,,\label{y^2selfinaldef}\\
\langle xy\rangle_s(t) & =0\;. \label{xyselfinaldef}
\end{eqnarray}
The salt patch is then exponentially stretched in the dilating direction ($x$) while it is compressed in the ($y$) direction toward the salt Batchelor scale $\ell_{B,s} = \sqrt{D_s/\sigma}$, which corresponds to a quasi-static equilibrium between compression and diffusion.

%%%%%%%
\subsection{Case of colloids, pure deformation and diffusiophoresis}

When released together with salt, colloids will have a drift velocity $\mathbf{v}_{\mathrm{dp}} = D_{\mathrm{dp}}\; \nabla\log S$ with respect to the flow $\mathbf{v} = (\sigma x, -\sigma y)$. 
As obtained from equations (\ref{gauss_s}), (\ref{x^2selfinaldef})-(\ref{xyselfinaldef}), the salt concentration is of the form
\begin{equation}
S(x,y,t)=A(t)\exp\left[-\frac{1}{2} \left(\frac{x^2}{\langle x^2\rangle_s(t)}+\frac{y^2}{\langle y^2\rangle_s(t)}\right)\right]\,,
\label{eq:champ_sel_def}
\end{equation}
so that the drift velocity, $\mathbf{v}_{\mathrm{dp}} = D_{\mathrm{dp}}\; \nabla\log S$, writes
\begin{equation}
\mathbf{v}_{\mathrm{dp}}(x,y,t) = D_{\mathrm{dp}}\; \left(\displaystyle\frac{-x}{\langle x^2\rangle_s(t)}, \displaystyle\frac{-y}{\langle y^2\rangle_s(t)}\right)\,.
\end{equation}
It is also a linear flow whose gradients are functions of time only so that the colloid concentration will also remain Gaussian at all times. 
Although $\mathbf{v}_{\mathrm{dp}}$ looks similar to $\mathbf{v}$, it has a non zero divergence $\nabla \cdot \mathbf{v}_{\mathrm{dp}} = -D_\mathrm{dp} (1/\langle x^2\rangle_s+1/\langle y^2\rangle_s)$ which is constant in space. 
This drift therefore acts as a compressing motion if $D_\mathrm{dp} > 0$ while it accelerates mixing if $D_\mathrm{dp} < 0$. 
Nevertheless $\iint_\infty C(x,y,t) dx dy=N_c$ is still a conserved quantity, and one can compute moments of the colloid concentration field as
\begin{equation}
\langle x^\alpha y^\beta\rangle_c(t)= \frac{1}{N_c} \iint_\infty x^\alpha y^\beta C(x,y,t) dx dy,
\end{equation}
to get the instantaneous colloid concentration field
\begin{equation}
C(x,y,t)=\frac{N_c}{2\pi \sqrt{\Delta_c(t)}}\,\exp\left(-\frac{1}{2\Delta_c(t)}\bigl[\langle y^2\rangle_c\,x^2-2\langle xy\rangle_c\, xy+\langle x^2\rangle_c\, y^2\bigr]\right).\label{gaussc}\\
\end{equation}
Applying the same procedure as in the previous section, second-order moments are solutions of the system:
\begin{eqnarray}
\frac{d\langle x^2\rangle_c}{dt}&=&- 2 D_{\mathrm{dp}}\frac{\langle x^2\rangle_c}{\langle x^2\rangle_s}+2\sigma \langle x^2\rangle_c+2D_c
\label{eq:x^2coldef}\\
\frac{d\langle xy\rangle_c}{dt}&=&-D_{\mathrm{dp}}\bigl[\langle x^2\rangle_s^{-1}+\langle y^2\rangle_s^{-1}\bigr]\;\langle xy\rangle_c\label{eq:xycoldef}\\
\frac{d\langle y^2\rangle_c}{dt}&=&- 2 D_{\mathrm{dp}}\frac{\langle y^2\rangle_c}{\langle y^2\rangle_s}-2\sigma \langle y^2\rangle_c+2D_c\:.
\label{eq:y^2coldef}
\end{eqnarray}
with initial conditions $\langle x^2\rangle_c(0)=x_{0,c}^2$, $\langle y^2\rangle_c(0)=y_{0,c}^2$, $\langle xy\rangle_c(0)=0$ so that $\langle xy\rangle_c(t)=0$. 
The solutions to ODEs for $\langle x^2\rangle_c$ and $\langle y^2\rangle_c$ can be obtained analytically and write: 
\begin{eqnarray}
\langle x^2\rangle_c(t)~&=&\left(x_{0,c}^2 -\frac{D_c}{D_{\mathrm{dp}}+D_s}x_{0,s}^2\right)\left(\frac{D_s}{x_{0,s}^2 \sigma}\bigl[1-\exp(-2\sigma t)\bigr]+1\right)^{-D_\mathrm{dp}/D_s}\!\!\!\!\!\!\!\!\!\exp(2\sigma t)\nonumber\\
&&\qquad+\frac{D_c}{D_{\mathrm{dp}}+D_s}x_{0,s}^2\,\left(\exp(2\sigma t)+\frac{D_s}{x_{0,s}^2 \sigma }[\exp(2\sigma t)-1]\right)\, , \label{eq:x^2colfinaldef}
\end{eqnarray}
%\begin{eqnarray}
%\langle xy\rangle_c(t)~&=&0\,
%\end{eqnarray}
\begin{eqnarray}
\langle y^2\rangle_c(t)~&=&\left(y_{0,c}^2 -\frac{D_c}{D_{\mathrm{dp}}+D_s}y_{0,s}^2\right)\left(\frac{D_s}{y_{0,s}^2\sigma}\bigl[\exp(2\sigma t)-1\bigr]+1\right)^{-D_\mathrm{dp}/D_s}\!\!\!\!\!\!\!\!\!\exp(-2\sigma t)\nonumber\\
&&\qquad+\frac{D_c}{D_{\mathrm{dp}}+D_s}y_{0,s}^2\,\left(\exp(-2\sigma t)+\frac{D_s}{y_{0,s}^2 \sigma}[1-\exp(-2\sigma t)]\right)\,.\label{eq:y^2colfinaldef}
\end{eqnarray} 

\unitlength=1.mm
\begin{figure}
\includegraphics{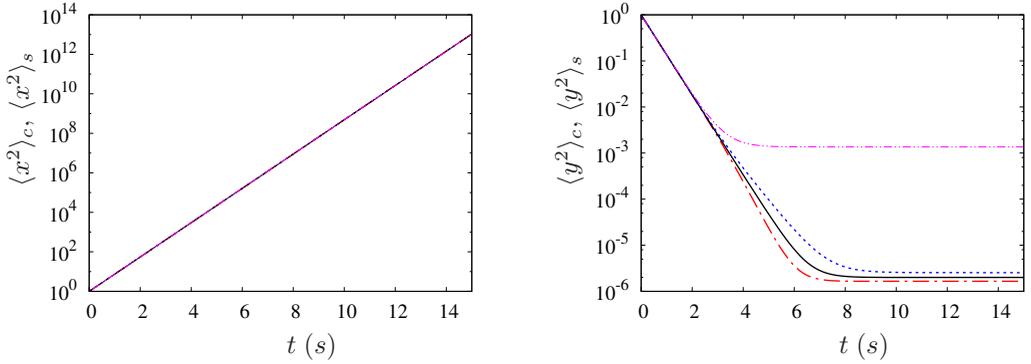}
\caption{
Time evolution of $\langle x^2\rangle_c$, $\langle x^2\rangle_s$ and $\langle y^2\rangle_c$, $\langle y^2\rangle_s$ in the case of a pure deformation. Parameters are similar to that of experiments ($D_s=1360\,\mu\mathrm{m^2s^{-1}}$, $D_c=2\,\mu\mathrm{m}^2s^{-1}$, $D_\mathrm{dp}=290\,\mu\mathrm{m^2\, s^{-1}}$), $y_{0,c}^2=x_{0,c}^2=1\,\mathrm{mm^2}$, and $\sigma=1\,\mathrm{s}^{-1}$, corresponding to $Pe_c=5.\,10^5$ and $Pe_s=735$. \textcolor{red}{\textbf{$-\cdot-\cdot-$}}: ``salt-attracting" case, $D_\mathrm{dp}=290\,\mu\mathrm{m}^2s^{-1}$; 
\textcolor{black}{---}: ``no-salt" case, $D_\mathrm{dp}=0\,\mu\mathrm{m^2\, s^{-1}}$;
\textcolor{blue}{- - -}: ``salt-repelling" case, $D_\mathrm{dp}=-290\,\mu\mathrm{m^2\, s^{-1}}$;
\textcolor{magenta}{\textbf{$-\!\cdot\!\cdot\!-\!\cdot\!\cdot\!-$}}: salt.
}
\label{fig:diffusio_deformation_D_dp}
\end{figure}
Figure \ref{fig:diffusio_deformation_D_dp} displays the time evolution of the lengths $\langle x^2\rangle_c(t)$ and $\langle y^2\rangle_c(t)$ for a set of parameters corresponding to those of the experiments, with $Pe_c=5.\,10^5$ and $Pe_s=735$. It shows that the evolution along the dilating direction, which corresponds to an exponential growth, is almost not affected either by diffusion or diffusiophoresis. 
On the other hand, we observe a clear influence of diffusiophoresis on the small scale, {\it e.g.} in the contracting direction $(y)$. 
This is especially true in the initial stage, before $\langle y^2\rangle_c$ reaches its final value $\langle y^2\rangle_c(\infty) = D_cD_s/(\sigma (D_\mathrm{dp} + D_s))$. 
As a consequence, the Batchelor scale is affected by diffusiophoresis so that the small scale becomes finer in the salt-attracting case ($D_\mathrm{dp} >0$) while it is coarser in the salt-repelling case ($D_\mathrm{dp} <0$), which is consistent with experimental results obtained in \cite{bib:Maugeretal2016}. The Batchelor scale can be computed as $\ell_{B,c} = \sqrt{D_\mathrm{eff}/\sigma}$, where $D_\mathrm{eff} = D_cD_s/(D_\mathrm{dp} + D_s)$. 
This relationship could serve as a definition for an effective diffusivity as was initially proposed in \cite{bib:Deseigneetal2014} to interpret the salt-repelling case, although it can be seen in equation (\ref{eq:y^2colfinaldef}) that the evolution of $\langle y^2\rangle_c(t)$ is not obtained by replacing $D_s$ by $D_\mathrm{eff}$ in equation (\ref{y^2selfinaldef}). 
Diffusiophoresis does not result in a process that can be considered as purely diffusive, with an effective diffusivity, as already stressed in \cite{bib:Volketal2014}.

%%%%%%%%%%%%%%%%%%%
\section{Dispersion in a shear flow}
\label{pureshear}

\subsection{Shear dispersion of salt}

The second problem we address is shear dispersion of Gaussian patches of salt and colloids under the action of the linear velocity field $\mathbf{v} = (\gamma y, 0)$ ($\gamma \geq 0$). We define again the P\'eclet numbers using the ($y$) direction for both salt $Pe_s = {\gamma y_{0,s}^2}/{D_s}$ and colloid $Pe_c = {\gamma y_{0,c}^2}/{D_c}$. \\

Applying the method of moments, one gets the coupled set of equations:
\begin{eqnarray}
\frac{d\langle x^2\rangle_s}{dt}&-2\gamma \langle xy\rangle_s &=\;2D_s
\label{eq:x^2sel_cisaillement}\\
\frac{d\langle xy\rangle_s}{dt}&-\gamma\langle y^2\rangle_s&=\;0\\
\frac{d\langle y^2\rangle_s}{dt}& &=\; 2D_s\;,
\label{eq:y^2sel_cisaillement}
\end{eqnarray}
which can be integrated with initial conditions $(\langle x^2\rangle_c(0),\langle y^2\rangle_c(0),\langle xy\rangle_c(0))=(x_{0,c}^2,y_{0,c}^2, 0)$ to obtain
\begin{eqnarray}
\langle y^2\rangle_s&=&y_{0,s}^2+2D_s t
\label{eq:y^2finalsel_cisaillement}\\
\langle xy\rangle_s&=&\gamma y_{0,s}^2 t+\gamma D_s t^2\label{eq:xyfinalsel_cisaillement}\\
\langle x^2\rangle_s &=& x_{0,s}^2+2D_s t+\gamma^2 y_{0,s}^2 t^2+\frac{2}{3}\gamma^2 D_s t^3.
\label{eq:x^2finalsel_cisaillement}
\end{eqnarray}

Inserting those functions in expression (\ref{gauss_s}), one recovers the Gaussian solution to this problem obtained by \cite{bib:okubo}. %As can be observed in equations (\ref{eq:y^2finalsel_cisaillement}), (\ref{eq:xyfinalsel_cisaillement}), (\ref{eq:x^2finalsel_cisaillement}), all characteristic lengths 
As can be observed, $\langle x^2\rangle_s$, $\langle xy\rangle_s$ and $\langle y^2\rangle_s$ are increasing functions of time with a diffusive growth in the ($y$) direction and a super diffusive growth in the ($x$) direction, although it could have been expected that the patch is compressed in some direction. This is due to the shear motion which tilts the patch toward the ($x$) direction as indicated by the growth of $\langle xy\rangle_s$. It is possible to define the variance of a large scale $\langle X^2\rangle_s$ and of a small scale $\langle Y^2\rangle_s$ by using the principal axes of the quadratic form $\langle y^2\rangle_s\,x^2-2\langle xy\rangle_s\, xy+\langle x^2\rangle_s\, y^2$. With these new axes, one has $\langle XY\rangle_s=0$, and:
\begin{eqnarray}
\langle X^2\rangle_s&=&\frac{2\Delta_s}{\langle x^2\rangle_s+\langle y^2\rangle_s-\sqrt{(\langle x^2\rangle_s-\langle y^2\rangle_s)^2+4\langle xy\rangle_s^2}}\label{eq:gauss_grande_echelle}\\
\langle Y^2\rangle_s&=&\frac{2\Delta_s}{\langle x^2\rangle_s+\langle y^2\rangle_s+\sqrt{(\langle x^2\rangle_s-\langle y^2\rangle_s)^2+4\langle xy\rangle_s^2}}\;,\label{eq:gauss_petite_echelle}
\end{eqnarray}
so that the area of the patch is $\mathcal{A}_s=\sqrt{\Delta_s}=\sqrt{\langle X^2\rangle_s \langle Y^2\rangle_s}$. As opposed to the previous case of pure deformation, the small scale does not converge toward a final value as shown in figure (\ref{fig:diffusio_shear_D_dp}). 
The small scale $\langle Y^2\rangle_s$ is first compressed as expected, but reaches a minimum value before slowly increasing again toward infinity. 
This is because here diffusion cannot be balanced in any direction as none of the ($x$) or ($y$) directions are dilating nor compressing as the eigenvalues of the velocity gradient matrix are zero.

%%%%%%%
\subsection{Shear dispersion of colloids under diffusiophoresis}

%\NOTE{we set X' for the colloids and X for the salt.}

Due to the action of shear which tilts the salt patch, the diffusiophoretic drift is now more complex. However $\mathbf{v}_{\mathrm{dp}}$ is still a linear velocity field as $S(x,y,t)$ is Gaussian, and involves gradients in all directions:
\begin{equation}
\mathbf{v}_{\mathrm{dp}} = \frac{D_{\mathrm{dp}}}{\Delta_s(t)}\;\left(\begin{array}{c}
-\langle y^2\rangle_s\,x+\langle xy\rangle_s\, y\\
\langle xy\rangle_s\, x-\langle x^2\rangle_s\,y
\end{array}\right) .
\end{equation}
Taking moments of equation (\ref{eqc}) with this more general velocity field, moments of the colloid concentration field are solutions of a system of fully coupled ODEs:
\begin{eqnarray}
&\!\!\!\!\!\!&\!\!\!\frac{d\langle x^2\rangle_c}{dt}-2\left(\gamma+\frac{D_{\mathrm{dp}}\langle xy\rangle_s}{\Delta_s(t)}\right) \langle xy\rangle_c+ 2\frac{D_{\mathrm{dp}} \langle y^2\rangle_s}{\Delta_s(t)} \langle x^2\rangle_c=2D_c\qquad\quad
\label{eq:x^2col_cisaillement}\\
&\!\!\!\!\!\!&\!\!\!\frac{d\langle xy\rangle_c}{dt}-\gamma\langle y^2\rangle_c -\frac{D_{\mathrm{dp}}\langle xy\rangle_s}{\Delta_s(t)}\bigl[\langle x^2\rangle_c+\langle y^2\rangle_c\bigr]+ \frac{D_{\mathrm{dp}}[\langle x^2\rangle_s+\langle y^2\rangle_s]}{\Delta_s(t)}\langle xy\rangle_c=0\qquad\quad\label{eq:xycol_cisaillement}\\
&\!\!\!\!\!\!&\!\!\!\frac{d\langle y^2\rangle_c}{dt} -2\frac{D_{\mathrm{dp}}\langle xy\rangle_s}{\Delta_s(t)}\langle xy\rangle_c+ 2\frac{D_{\mathrm{dp}} \langle x^2\rangle_s}{\Delta_s(t)} \langle y^2\rangle_c=2D_c%\qquad\quad
\label{eq:y^2col_cisaillement}
\end{eqnarray}
with same initial conditions as in the previous case. Such system has no known analytical solutions but can be integrated numerically for any set of parameters using standard techniques (fourth-order Runge-Kutta in the present case). 
As already observed for the salt, the colloid patch is expected to be tilted toward ($x$) axis so that we introduce its principal axes and compute 
\begin{eqnarray}
\langle X'^2\rangle_c&=&\frac{2\Delta_c}{\langle x^2\rangle_c+\langle y^2\rangle_c-\sqrt{(\langle x^2\rangle_c-\langle y^2\rangle_c)^2+4\langle xy\rangle_c^2}}\label{eq:gauss_grande_echelle_c}\\
\langle Y'^2\rangle_c&=&\frac{2\Delta_c}{\langle x^2\rangle_c+\langle y^2\rangle_c+\sqrt{(\langle x^2\rangle_c-\langle y^2\rangle_c)^2+4\langle xy\rangle_c^2}}\;.\label{eq:gauss_petite_echelle_c}
\end{eqnarray}
\begin{figure}%[h]
\includegraphics{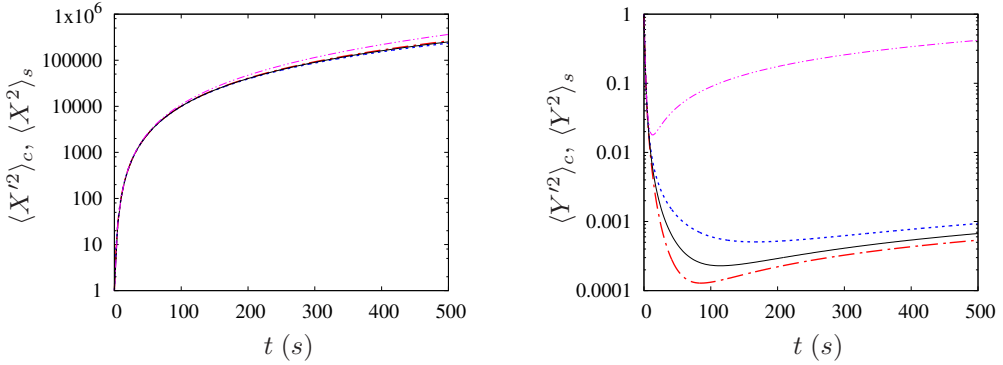}
\caption{
Time evolution of $\langle X'^2\rangle_c$, $\langle X^2\rangle_s$ and $\langle Y'^2\rangle_c$, $\langle Y^2\rangle_s$ for $\gamma=1\,\mathrm{s}^{-1}$. Parameters are the same as in the experiments ($D_s=1360\,\mu\mathrm{m^2s^{-1}}$, $D_c=2\mu\mathrm{m}^2s^{-1}$ and $D_\mathrm{dp}=290\mu\mathrm{m^2s^{-1}}$), $y_{0,c}^2=x_{0,c}^2=1\,\mathrm{mm^2}$, corresponding to $Pe_c=5.\,10^5$ and $Pe_s=735$.
\textcolor{red}{\textbf{$-\cdot-\cdot-$}}: ``salt-attracting" case; 
\textcolor{black}{---}: ``no-salt" case, $D_\mathrm{dp}=0\,\mu\mathrm{m^2\, s^{-1}}$;
\textcolor{blue}{- - -}: ``salt-repelling" case, $D_\mathrm{dp}=-290\,\mu\mathrm{m^2\, s^{-1}}$;
\textcolor{magenta}{\textbf{$-\!\cdot\!\cdot\!-\!\cdot\!\cdot\!-$}}: evolution for the salt.
}
\label{fig:diffusio_shear_D_dp}
\end{figure}
Note that the axes for the salt and colloids do not coincide so that the small and large scales are not measured exactly along the same direction. Figure \ref{fig:diffusio_shear_D_dp} displays the time evolution of the lengths $\langle X'^2\rangle_c(t)$ and $\langle Y'^2\rangle_c(t)$ for a set of parameters corresponding to those of the experiments, with $Pe_c=5.\,10^5$ and $Pe_s=735$. As already observed in the case of compression, we find that diffusiophoresis mostly affects the small scale which again becomes finer in the salt-attracting case ($D_\mathrm{dp} >0$) while it is coarser in the salt-repelling case ($D_\mathrm{dp} <0$), the effect being slightly larger in this second configuration.

%%%%%%%%%%%%%%%%%%%%%%%%%%%%%%%
\section{Mixing and compressibility}
\label{mixcompress}
\begin{figure}%[h]
\includegraphics{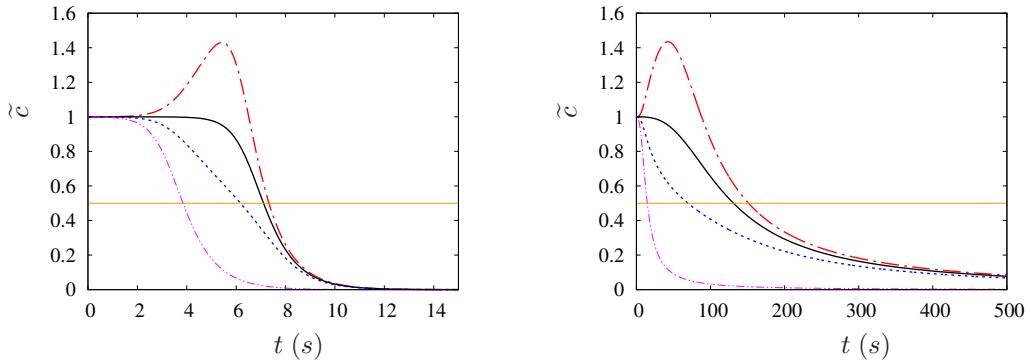}
\caption{
Time evolution of $\tilde c=\sqrt{\Delta_c(0)/\Delta_c}$ in the case of pure deformation (left) and pure shear (right). Same set of parameters as in experiments ($D_s=1360\,\mu\mathrm{m^2s^{-1}}$ and $D_c=2\,\mu\mathrm{m}^2s^{-1}$, $D_\mathrm{dp}=290\,\mu\mathrm{m}^2s^{-1}$), $y_{0,c}^2=x_{0,c}^2=1\,\mathrm{mm^2}$. Pure deformation case, $\sigma=1\,\mathrm{s}^{-1}$, $Pe_c=5.\,10^5$. Pure shear $\gamma=1\,\mathrm{s}^{-1}$, $Pe_c=5.\,10^5$. Legends are \textcolor{red}{\textbf{$-\cdot-\cdot-$}}: ``salt-attracting" case, $D_\mathrm{dp}=290\,\mu\mathrm{m}^2s^{-1}$; 
\textcolor{black}{---}: ``no-salt" case, $D_\mathrm{dp}=0\,\mu\mathrm{m^2\, s^{-1}}$;
\textcolor{blue}{- - -}: ``salt-repelling" case, $D_\mathrm{dp}=-290\,\mu\mathrm{m^2\, s^{-1}}$;
\textcolor{magenta}{\textbf{$-\!\cdot\!\cdot\!-\!\cdot\!\cdot\!-$}}: case of salt.
The horizontal line $y=1/2$ serves as measuring the mixing time, $T_\mathrm{mix}$, such that $\tilde c (T_\mathrm{mix}) = 1/2$.
}
\label{fig:evolution_c}
\end{figure}

%%%%%%%
\subsection{Compressibility and its link to changes in colloid concentration}

In the two previous cases, we found that both salt and colloid concentration profiles remain Gaussian with areas $\mathcal{A}_s=\sqrt{\Delta_s}$ and $\mathcal{A}_c=\sqrt{\Delta_c}$ changing as functions of time. Because the total colloid and salt content are conserved, the areas of the patches can be used as a measure of the respective concentrations ${s}(t)=N_s/\sqrt{\Delta_s}$ and ${c}(t)=N_c/\sqrt{\Delta_c}$. 

As all equations are linear, we will focus on the non-dimensional quantities $\tilde{s}(t)=\sqrt{\Delta_s(0)/\Delta_s(t)}$ and $\tilde{c}(t)=\sqrt{\Delta_c(0)/\Delta_c(t)}$. Figure \ref{fig:evolution_c} displays the time evolution of the concentrations $\tilde{c}(t)$ and $\tilde{s}(t)$ for the two cases of pure deformation (left) and pure shear (right), with the very same set of parameters and $\sigma=\gamma=1$ s$^{-1}$. As already observed in chaotic advection (\cite{bib:Deseigneetal2014,bib:Volketal2014}), the concentration decays much faster in the salt-repelling configuration (enhanced mixing), while it first increases toward a maximum in the salt-attracting configuration, resulting in a delayed mixing. This is here a direct consequence of the patches aspect ratio evolution which is strongly modified by diffusiophoresis at short time. All these changes can be interpreted by examining the area dilation factor, $\chi$, defined as:
\begin{equation}
\chi=\frac{1}{\mathcal{A}_c}\,\frac{d\mathcal{A}_c}{dt} = -\frac{d\ln\bigl(\tilde c(t)\bigr)}{dt} = \frac{1}{2 \Delta_c}\,\frac{d\Delta_c}{dt} \label{eq:chi_c_tilde}.%,
\end{equation}
This equation establishes a direct link between the evolution of $\tilde c$ and compressible effects as pointed out in \cite{bib:Volketal2014}. Indeed, using second-order moments for both salt and colloids $\Delta_c(t)=\langle x^2\rangle_c\langle y^2\rangle_c-\langle xy\rangle_c^2$, one obtains the general expression:
\begin{eqnarray}
\chi&=&\frac{1}{2\Delta_c(t)}\,\frac{d\Delta_c(t)}{dt} \label{eq:chidelta}\\
&=&\frac{1}{2\Delta_c}\left(\frac{d\langle x^2\rangle_c}{dt}\langle y^2\rangle_c+\langle x^2\rangle_c\frac{d\langle y^2\rangle_c}{dt}-2\langle xy\rangle_c\frac{d\langle xy\rangle_c}{dt}\right)
\end{eqnarray}
{\bf In the case of pure deformation,} the cross-term vanishes. Using equations (\ref{eq:x^2coldef}) and (\ref{eq:y^2coldef}), the area dilation can be rewritten:
\begin{equation}
\chi= D_c\left(\frac{1}{\langle x^2\rangle_c}+\frac{1}{\langle y^2\rangle_c}\right) -D_\mathrm{dp}\left(\frac{1}{\langle x^2\rangle_s}+\frac{1}{\langle y^2\rangle_s}\right).
%-\frac{d\ln\bigl(\tilde c(t)\bigr)}{dt} =
\label{eq:chi_deformation_pure}
\end{equation}
Equation (\ref{eq:chi_deformation_pure}) shows that the area of the colloid patch varies as a sum of the diffusion, which tends to make it grow in size, and diffusiophoresis: 
when $D_\mathrm{dp} >0$ (salt-attracting configuration), the area first contracts because here $D_c \ll |D_\mathrm{dp}|$, while the patch spreads faster when $D_\mathrm{dp} < 0$ (salt-repelling configuration). 
It is interesting to note that the diffusiophoretic contribution to the dilation factor is exactly $\nabla \cdot \mathbf{v}_{\mathrm{dp}}$, in agreement with the colloid concentration budget established in \cite{bib:Volketal2014}, or with more general results for inertial particles (\cite{bib:Metcalfeetal2012}).

\noindent {\bf In the case of shear dispersion,} the cross-term does not vanish as both salt and colloid patches are tilted by the flow. Using equations  (\ref{eq:x^2col_cisaillement}), (\ref{eq:xycol_cisaillement}) and (\ref{eq:y^2col_cisaillement}),  the area dilation has the more complex form:

\begin{equation}
\chi =D_c\,\frac{\langle x^2\rangle_c+\langle y^2\rangle_c}{\Delta_c} -D_\mathrm{dp}\,\frac{\langle x^2\rangle_s+\langle y^2\rangle_s}{\Delta_s}
\end{equation}

Introducing the two systems of principal axes for the colloids $(X',Y')$, and for the salt $(X,Y)$, one can obtain a similar expression as in the case of pure deformation. The equation writes: 
\begin{equation}
\chi=D_c\,\left(\frac{1}{\langle X'^2\rangle_c}+\frac{1}{\langle Y'^2\rangle_c}\right) -D_\mathrm{dp}\,\left(\frac{1}{\langle X^2\rangle_s}+\frac{1}{\langle Y^2\rangle_s}\right) \,,
\label{eq:dilatation_axes principaux}
\end{equation}
from which it is visible that the second term is again exactly $\nabla \cdot \mathbf{v}_\mathrm{dp}$. This shows that the interpretation of diffusiophoresis in terms of a competition/cooperation between diffusion and compressible effects is very robust and general. In both cases we find that the flow parameters do not enter the result explicitly because $\mathbf{v}$ is divergence free. The effect of the velocity field is in fact hidden and it is only when investigating the evolution of the various length scales that its properties are directly visible. \\

%%%%%%%
\subsection{Evolution of the mixing time in the limit $D_c \rightarrow 0$} 

We now investigate how much time is needed to mix the patches in the different cases. 
From the previous section, it may be thought that mixing processes in both flows are similar because the two graphics displayed in figure \ref{fig:evolution_c} look similar on first inspection. 
However, because of exponential stretching in the pure deformation versus algebraic stretching in the pure shear, the mixing efficiencies of these two flows are in fact very different, although they correspond to the same P\'eclet number $Pe_c = 5 ~10^5$. 
This can be quantified by the mixing time, $T_\mathrm{mix}$, defined as the time when the relative concentration is divided by $2$: $\tilde c (T_\mathrm{mix}) = 1/2$. From figure \ref{fig:evolution_c}, we have $\sigma T_\mathrm{mix}(\mathrm{deformation}) \simeq 6.5$, which is one order of magnitude smaller than $\gamma T_\mathrm{mix}(\mathrm{shear}) \simeq 150$. 

\begin{figure}%[h]
\includegraphics{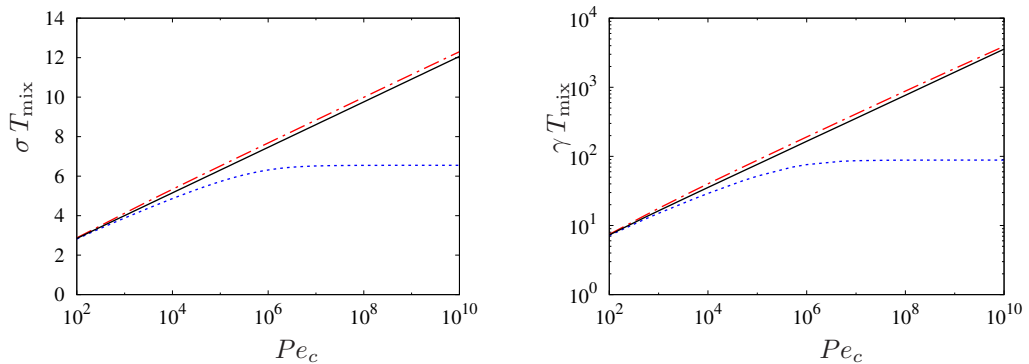}
\caption{
Evolution of the mixing time $T_\mathrm{mix}$ in the case of pure deformation (left, $\sigma=1\,\mathrm{s}^{-1}$) and pure shear (right, $\gamma=1\,\mathrm{s}^{-1}$) for a fixed shear rate and variable colloid diffusion coefficient $D_c \in [0.2, 2 \cdot 10^{5}]\mu\mathrm{m}^2s^{-1}$. Same set of parameters as in experiments $D_s=1360\,\mu\mathrm{m^2s^{-1}}$, $D_\mathrm{dp}=290\,\mu\mathrm{m}^2s^{-1}$, $y_{0,c}^2=x_{0,c}^2=1\,\mathrm{mm^2}$. Legends are \textcolor{red}{\textbf{$-\cdot-\cdot-$}}: ``salt-attracting" case, $D_\mathrm{dp}=290\,\mu\mathrm{m}^2s^{-1}$; 
\textcolor{black}{---}: ``no-salt" case, $D_\mathrm{dp}=0\,\mu\mathrm{m^2\, s^{-1}}$;
\textcolor{blue}{- - -}: ``salt-repelling" case, $D_\mathrm{dp}=-290\,\mu\mathrm{m^2\, s^{-1}}$.
}
\label{fig:mixing_time}
\end{figure}

Figure \ref{fig:mixing_time} displays the evolution of $T_\mathrm{mix}$ in both cases when increasing $Pe_c$ by decreasing the colloid diffusion coefficient $D_c$ with all other parameters maintained fixed so that the salt P\'eclet number remains $Pe_s=735$\footnote{Parameters are $D_s=1360\,\mu\mathrm{m^2s^{-1}}$, $D_\mathrm{dp}=290\,\mu\mathrm{m}^2s^{-1}$, $y_{0,c}^2=x_{0,c}^2=1\,\mathrm{mm^2}$, the shear rate being $\sigma=1$ $\mathrm{s^{-1}}$ (deformation) and $\gamma = 1$ $\mathrm{s^{-1}}$ (shear).}. 
In the case of pure deformation, $T_\mathrm{mix}$ increases logarithmically with $Pe_c$, and we find that the ratio of the mixing times with and without diffusiophoresis is nearly constant at moderate P\'eclet number $Pe_c \leq 10^5$. 
Such a result can be explained by recalling that the large scale is weakly affected by diffusiophoresis and expands exponentially ${\langle x^2 \rangle_c} \sim x_{0,c}^2 \exp{2 \sigma t}$ while the small scale is compressed toward the modified Batchelor scale ${\langle y^2 \rangle_c} = \ell_{B,c}^2= {D_c D_{s}}/{(D_\mathrm{dp} + D_s)\gamma}$. 
Defining $T_\mathrm{mix}$ as the time needed for the concentration to be half of its initial value, one obtains:
\begin{eqnarray}
\frac{\langle x^2\rangle_c \langle y^2\rangle_c}{x_{0,c}^2y_{0,c}^2}&=&4, \hspace{0.5cm} t=T_\mathrm{mix} \\
T_\mathrm{mix}&=&\frac{1}{2\sigma}\ln\left(4 Pe_c\,\frac{D_\mathrm{dp}+D_s}{D_s}\right)\;.
\end{eqnarray}
which gives a qualitative picture of how the $T_\mathrm{mix}$ is affected by diffusiophoresis at moderate $Pe_c$. 

In the case of pure shear, $T_\mathrm{mix}$ is found to increase as a power law $T_\mathrm{mix} \propto Pe_c^{1/3}/\gamma$. Such algebraic scaling is  a direct consequence of equation (\ref{eq:x^2finalsel_cisaillement}), and is typical of shear dispersion (\cite{bib:younggarrett1982,bib:rhines_young1983}). It is obtained by assuming that the patch grows as $\langle x^2 \rangle_c \sim D_c \gamma^2 t^3$ in the $x$ direction so that it has doubled in size in a time $T_\mathrm{mix}$ (here $x_{0,c}^2=y_{0,c}^2$). In this second case too, we find that in the moderate P\'eclet number range the mixing time is always slightly larger in the salt-attracting configuration ($D_\mathrm{dp}>0$) whereas $T_\mathrm{mix}$ is smaller in the salt-repelling configuration ($D_\mathrm{dp} <0$). These observations are consistent  with previous experimental and numerical studies of the mixing time (\cite{bib:Deseigneetal2014,bib:Volketal2014}), especially in the case of pure deformation for which the mixing time presents the same logarithmic scaling as in the chaotic regime due to the action of compression. 

In the high P\'eclet number range, we observe that $T_\mathrm{mix}$ saturates (close to $\sigma T_\mathrm{mix}(\mathrm{deformation)} = 6.5$ and $\gamma T_\mathrm{mix}(\mathrm{shear)} = 88$ respectively). 
This is a very interesting behaviour because it shows that the colloids can be mixed efficiently although they have a vanishingly small diffusion coefficient. 
Such a property is uncommon, and is not explained by the two qualitative pictures given previously. 
\begin{figure}%[h]
\includegraphics{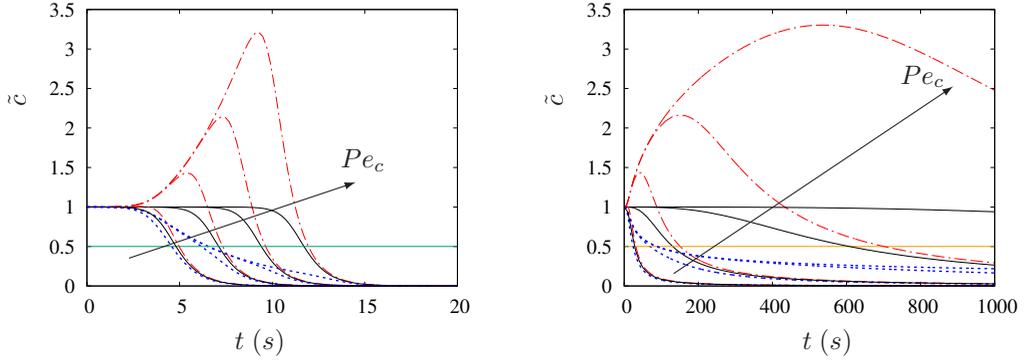}
\caption{
Time evolution of $\tilde c=\sqrt{\Delta_c(0)/\Delta_c}$ in the case of pure deformation (left, $\sigma=1\,\mathrm{s}^{-1}$) and pure shear (right, $\gamma=1\,\mathrm{s}^{-1}$) for a fixed shear rate and four values of the colloid diffusion coefficient $D_c=[0.02, 2, 200, 20000]\mu\mathrm{m}^2s^{-1}$. Same set of parameters as in experiments $D_s=1360\,\mu\mathrm{m^2s^{-1}}$, $D_\mathrm{dp}=290\,\mu\mathrm{m}^2s^{-1}$, $y_{0,c}^2=x_{0,c}^2=1\,\mathrm{mm^2}$. Legends are \textcolor{red}{\textbf{$-\cdot-\cdot-$}}: ``salt-attracting" case, $D_\mathrm{dp}=290\,\mu\mathrm{m}^2s^{-1}$; 
\textcolor{black}{---}: ``no-salt" case, $D_\mathrm{dp}=0\,\mu\mathrm{m^2\, s^{-1}}$;
\textcolor{blue}{- - -}: ``salt-repelling" case, $D_\mathrm{dp}=-290\,\mu\mathrm{m^2\, s^{-1}}$.
}
\label{fig:evolutionc_Dc}
\end{figure}
In order to understand this behaviour we plot the evolution $\tilde c$ in figure \ref{fig:evolutionc_Dc} for four values of the P\'eclet number $Pe_c=5\,10^3,  5\,10^5, 5\,10^7, 5\,10^9$. From figure \ref{fig:evolutionc_Dc}, we observe that increasing the P\'eclet number results in a shift of $\tilde c(t)$ toward larger times both for the reference and the salt-attracting configuration so that $T_\mathrm{mix}$ increases at increasing $Pe_c$. On the opposite, no such shift is observed in the salt-repelling configuration for which increasing the P\'eclet number no longer changes the evolution of $\tilde c(t)$ at short time as soon as $Pe_c \geq 10^5$, and it is only on a longer time scale that differences between the curves can be observed. We thus conclude that the saturation values reported above strongly depends on the precise definition of $T_\mathrm{mix}$. Indeed, defining $T_\mathrm{mix}$ as the time needed to divide the concentration by $10$ ($90\%$ decrease) would extend the range of P\'eclet number in which we observe an increase, with larger saturation values. If this shows that the present result is robust, it points out the difficulty of defining a mixing efficiency through a single quantity, such as $T_\mathrm{mix}$, as soon as the curves change their shape when varying the parameters.

\begin{figure}%[h]
\includegraphics{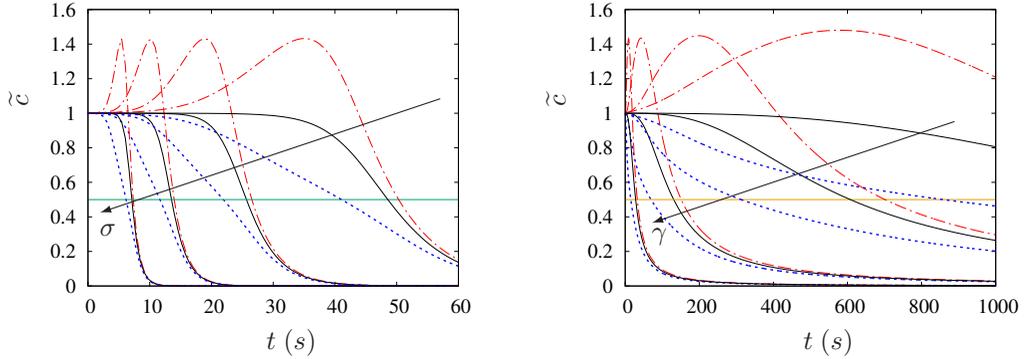}
\caption{
Time evolution of $\tilde c=\sqrt{\Delta_c(0)/\Delta_c}$ for a fixed colloid diffusion coefficient and different values of the shear. Left, pure deformation $\sigma=1\,\mathrm{s}^{-1}$, $0.5\,\mathrm{s}^{-1}$, $0.25\,\mathrm{s}^{-1}$, $0.125\,\mathrm{s}^{-1}$; right, pure shear $\gamma=0.02\,\mathrm{s}^{-1}$, $0.1\,\mathrm{s}^{-1}$, $1\,\mathrm{s}^{-1}$, $10\,\mathrm{s}^{-1}$. Same set of parameters as in experiments ($D_s=1360\,\mu\mathrm{m^2s^{-1}}$, $D_c=2\mu\mathrm{m}^2s^{-1}$, $D_\mathrm{dp}=290\,\mu\mathrm{m}^2s^{-1}$), $y_{0,c}^2=x_{0,c}^2=1\,\mathrm{mm^2}$. Legends are \textcolor{red}{\textbf{$-\cdot-\cdot-$}}: ``salt-attracting" case, $D_\mathrm{dp}=290\,\mu\mathrm{m}^2s^{-1}$; 
\textcolor{black}{---}: ``no-salt" case, $D_\mathrm{dp}=0\,\mu\mathrm{m^2\, s^{-1}}$;
\textcolor{blue}{- - -}: ``salt-repelling" case, $D_\mathrm{dp}=-290\,\mu\mathrm{m^2\, s^{-1}}$.
}
\label{fig:evolutionc_gama}
\end{figure}

\subsection{The maximum concentration does not depend on flow properties ($D_\mathrm{dp}>0$).}

When investigating the influence of the colloid diffusivity $D_c$ in the salt-attracting case (figure \ref{fig:evolutionc_Dc}), it appears that $\tilde c(t)$ reaches a maximum value which increases when decreasing $D_c$. Moreover this maximum of concentration seems not to depend directly on the flow. Indeed, the maximum $\tilde{c}_\mathrm{max}$ is found to be nearly the same in the two cases of deformation and shear displayed in figure \ref{fig:evolutionc_Dc} (see also figure \ref{fig:evolution_c}). Moreover, all parameters the same except $\gamma$ or $\sigma$, we observe in figure \ref{fig:evolutionc_gama} that $\tilde{c}_\mathrm{max}$ is insensitive to the stirring properties of the flow. In this large P\'eclet number range $Pe_c \geq 10^3$, $\tilde{c}_\mathrm{max}$ seems to depend only on chemical properties and initial lengths.\\

A qualitative analysis developed below shows that in the case of initially round patches, the maximum of colloid concentration is well predicted by the relation:
\begin{equation}
%\boxed{
\tilde{c}_\mathrm{max}=\displaystyle \left(\frac{D_\mathrm{dp}}{D_c}\frac{\langle Y'^2\rangle_c(0)}{\langle Y^2\rangle_s(0)}\right)^{\frac{D_\mathrm{dp}}{2(D_\mathrm{dp}+D_s)}}\,.
\label{eq:c_tilde_max2}
\end{equation}

In order to establish this relation, we first start with equations (\ref{eq:chidelta}) and (\ref{eq:dilatation_axes principaux}) which describe the evolution of both salt and colloids compressibilities, and rewrite them as:
\begin{eqnarray}
\frac{1}{2\Delta_s}\,\frac{d\Delta_s}{dt} & = & {D_s}\left(\frac{1}{\langle X^2\rangle_s}+\frac{1}{\langle Y^2\rangle_s}\right)\\
\frac{1}{2\Delta_c}\,\frac{d\Delta_c}{dt} & = & -{D_\mathrm{dp}}\left(\frac{1}{\langle X^2\rangle_s} + \frac{1}{\langle Y^2\rangle_s}\right) + {D_{c}}\left(\frac{1}{\langle X'^2\rangle_c} + \frac{1}{\langle Y'^2\rangle_c}\right), \label{eq:deltac}
\end{eqnarray}
where we used the principal axes of both Gaussian distributions. 
In the initial stage of colloid mixing, diffusive effects are negligible as compared to diffusiophoretic effect because $D_c \ll D_\mathrm{dp}$, so that
\begin{eqnarray}
\frac{1}{2\Delta_c}\,\frac{d\Delta_c}{dt} & \approx & -{D_\mathrm{dp}}\left(\frac{1}{\langle X^2\rangle_s} + \frac{1}{\langle Y^2\rangle_s}\right).
\end{eqnarray}
In this first stage, the area of both species are related to each other by the relation
\begin{equation}
\frac{1}{D_\mathrm{dp}}\,\ln\frac{\Delta_c(t)}{\Delta_c(0)}+\frac{1}{D_s}\,\ln\frac{\Delta_s(t)}{\Delta_s(0)} \approx 0 , 
\label{eq:eq1_Delta_c_Delta_s}
\end{equation}
which explains why all curves for the salt-attracting configuration follow the same initial evolution in figure \ref{fig:evolutionc_Dc} where only $D_c$ is varied. We will assume that this relation holds for $t\leq t^\star$ such that $\tilde{c}(t^\star)=\tilde{c}_\mathrm{max}$, which is an approximation as the role of $D_c$ is neglected and may lead to a slight overestimation of $\tilde{c}_\mathrm{max}$. When $\tilde{c}$ reaches its maximum value, diffusive and diffusiophoretic effects compensate so that $\Delta_c$ reaches a minimum and its time derivative is zero. As the small scales of both species are much smaller than the corresponding large scales, $\langle Y^2\rangle_s(t^\star)$ and $\langle Y'^2\rangle_c(t^\star)$ are then linked by the relation (equation (\ref{eq:deltac})):
\begin{equation}
\langle Y'^2\rangle_c(t^\star)=\frac{D_{c}}{D_\mathrm{dp}}\langle Y^2\rangle_s(t^\star)
\label{eq:Y2_s_Y2_c}
\end{equation}

In the case of initially round patches, the large scales of both species are only weakly affected by diffusion and diffusiophoresis so that $\langle X'^2 \rangle_c(t)/\langle X'^2 \rangle_c(0)  \approx \langle X^2 \rangle_s(t)/\langle X^2 \rangle_s(0)$ at any time. 
This leads to the relation:
\begin{equation}
\Delta_c(t^\star)=\frac{D_{c}}{D_\mathrm{dp}} \frac{\langle X'^2\rangle_c(0)}{\langle X^2\rangle_s(0)}  \Delta_s(t^\star), 
\label{eq:Deltasc}
\end{equation}
which we chose to write
\begin{equation}
\frac{\Delta_s(t^\star)}{\Delta_s(0)}=\frac{\Delta_c(t^\star)}{\Delta_c(0)} \frac{D_\mathrm{dp}}{D_{c}} \frac{\langle Y'^2\rangle_c(0)}{\langle Y^2\rangle_s(0)}  . %\hspace{1cm} \frac{\Delta_c(0)}{\Delta_s(0)} = \frac{\langle Y'^2\rangle_c(0)}{\langle Y^2\rangle_s(0)},
\end{equation}

This last relation can be combined with equation (\ref{eq:eq1_Delta_c_Delta_s}) to get the formula given in equation (\ref{eq:c_tilde_max2}). 
Note that in the case of pure deformation $x$ and $y$ directions decouple so that equation (\ref{eq:c_tilde_max2}) holds also for non-round patches ({\it i.e.} with arbitrary initial aspect ratios). 
Note also that we used the hypothesis that the distributions of salt and colloids possess a large and a small scale, so that equation (\ref{eq:c_tilde_max2}) would not hold in the absence of flow (pure diffusiophoresis).

\begin{figure}%[h]
\begin{center}
%{\includegraphics[width=54mm]{figure7.eps}}
\includegraphics{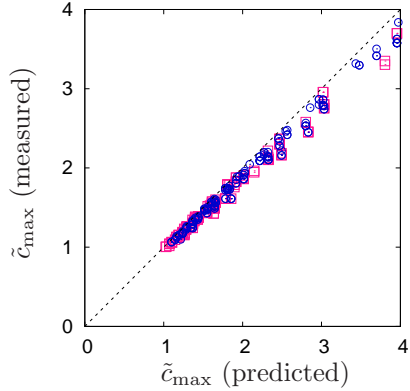}
\end{center}
\caption{Evolution of the relative maximum measured from $\tilde c(t)$ as a function of the predicted value using equation (\ref{eq:c_tilde_max2}). Legends are  (\textcolor{blue}{\textbf{$\circ$}}): pure shear. (\textcolor{red}{\textbf{$\square$}}): pure deformation. (\textcolor{black}{-~-}): straight line of equation $y=x$.}
\label{fig:cthcmes}
\end{figure}

The prediction of $\tilde{c}_\text{max}$ has been tested with initially round patches in both cases of pure shear and deformation by varying independently all parameters ($D_s$, $D_\mathrm{dp}$, $D_c$, $x_{0,s}=y_{0,s}$, $x_{0,c}=y_{0,c}$) over two decades with fixed values $\sigma=1$ s$^{-1}$ and $\gamma=1$ s$^{-1}$. 
Figure \ref{fig:cthcmes} displays the maximum concentration measured from $\tilde c(t)$ as a function of the predicted value. 
It is remarkable that the maximum is very well predicted by the relation (\ref{eq:c_tilde_max2}) with less than $10\%$ error. 
This shows that the hypothesis of taking into account colloid diffusion only very close to the maximum is a good approximation. 
As seen from this figure is an other striking result: by setting the experimental configuration of a small salt patch ($y_{0,s} \approx 1$ mm) and a larger colloid patch ($y_{0,c} \approx 10$ mm), it is in principle possible to increase the colloid concentration by a factor larger than $10$, which would correspond to strong demixing of the colloids.

\section{Conclusion}
\label{concl}

We have studied analytically the dispersion of patches of salt and colloids in linear velocity fields. The velocity fields (pure deformation and pure shear) were chosen to correspond to fundamental examples at the heart of our understanding of mixing (\cite{bib:bakuninbook}). Assuming the patches were initially released with Gaussian profiles, both cases could be analytically solved almost entirely so that results could be obtained for any set of parameters. 

An analytical solution was obtained in the case of pure deformation, which showed that diffusiophoresis led to a modification of the Batchelor scale $\ell_{B,c} = \sqrt{D_cD_s/\sigma (D_\mathrm{dp} + D_s)}$. The case of pure shear was found to be more complex, but equations for the evolutions of colloid concentration moments were obtained and solved numerically for various sets of parameters. 

Using the area of the patches $\mathcal{A}$, the evolution of the colloid global concentration (rescaled by its initial value) was studied. 
A prediction for the time, $T_\mathrm{mix}$, needed to decrease the concentration by a factor 2 was obtained. 
In both cases, we found that in the moderate P\'eclet number range the mixing time is always slightly larger in the salt-attracting configuration ($D_\mathrm{dp}>0$) whereas $T_\mathrm{mix}$ is smaller in the salt-repelling configuration ($D_\mathrm{dp} <0$). These observations are consistent  with previous experimental and numerical studies of the mixing time (\cite{bib:Deseigneetal2014,bib:Volketal2014}), especially in the case of pure deformation for which the mixing time presents the same logarithmic scaling as in the chaotic regime due to the action of compression. 

In all cases, the evolution of the concentration was intriguingly similar to the one observed experimentally in chaotic mixing, presenting a maximum in the salt-attracting configuration as the colloid concentration is first amplified before being attenuated. 
Using arguments based on compressibility, it was possible to obtain a prediction of the maximum concentration when dealing with initially round patches. 
%\begin{equation}
%\tilde{c}_\text{max}=\displaystyle \left(\frac{D_\mathrm{dp}}{D_c}\right)^{\frac{D_\mathrm{dp}}{2(D_\mathrm{dp}+D_s)}},
%\end{equation}
This prediction involves the existence of a flow, but does not involve the flow parameters, and compares very well with numerical and analytical results. 
Finally, we stress that while established in the case of linear flows, it also gives the correct order of magnitude (less than $3\%$ error) in the case of diffusiophoresis with chaotic advection in our numerical paper (\cite{bib:Volketal2014}). 
This points out that diffusiophoretic effects are not changed when increasing stirring. This observation also holds for the mixing time as we found that the ratio $T_\mathrm{mix}(D_\mathrm{dp}\neq 0)/T_\mathrm{mix}(D_\mathrm{dp} = 0)$ does not depend on the shear rate.\\

{\bf Acknowledgments} This collaborative work was supported by the French research programs ANR-16-CE30-0028 and 
LABEX iMUST (ANR-10-LABX-0064) of Universit\'e de Lyon, within the 
program``Investissements d'Avenir" (ANR-11-IDEX-0007) operated by the French National Research
Agency (ANR).

\bibliography{main}
\bibliographystyle{jfm}

\end{document}